\documentclass[twocolumn,preprintnumbers,floatfix]{revtex4-1}
\usepackage{amsmath}
\usepackage{amsfonts}
\usepackage{amssymb}
\usepackage{graphicx}
\def\be{\begin{equation}}
\def\ee{\end{equation}}
\def\bes{\begin{equation}\begin{split}&}
\def\ees{\end{split}\end{equation}}
\def\bi{\bibitem}

\begin{document}
\title{Validating variational principle for higher order theory of gravity.}
\author{Soumendranath Ruz}
 \email{ruzfromju@gmail.com}
 \affiliation {Dept.of Physics, University of Kalyani, Nadia, India - 741235}
\author{Kaushik Sarkar}
 \email{sarkarkaushik.rng@gmail.com}
 \affiliation {Dept.of Physics, University of Kalyani, Nadia, India - 741235}
\author{Nayem Sk.}
 \email{nayemsk1981@gmail.com}
 \affiliation {Dept.of Physics, University of Kalyani, Nadia, India - 741235}
\author{Abhik Kumar Sanyal}
 \email{sanyal\_ ak@yahoo.com}
 \affiliation {Dept.of Physics, Jangipur College, Murshidabad, India - 742213}

\begin{abstract}
\noindent
Metric variation of higher order theory of gravity requires to fix the Ricci scalar in addition to the metric tensor at the boundary. Fixing Ricci scalar at the boundary implies that the classical solutions are fixed once and forever to the de-Sitter or anti de-Sitter solutions. Here, we justify such requirement from the standpoint of Noether Symmetry.
\end{abstract}

\maketitle

It had been shown long ago \cite{Bunch} that under metric variation, no higher order term other than Gauss-Bonnet combination, can produce a suitable surface term under the only condition $\delta g_{\mu\nu}\Big{|}_{\partial\mathcal V} = 0$ at the boundary. If we concentrate upon $f(R)$ theory of gravity, which is a strong contender to an alternative to the dark energy, then under metric variation, the action
\be A = \int \sqrt {-g} \mathrm{d^4x} f(R)\ee
produces certain boundary term as
\bes \label{c5} \delta A = \int \mathrm{d^4x}\sqrt{-g}\left[ \Big(R_{\mu\nu} + g_{\mu\nu}\Box  - \nabla_\mu\nabla_\nu\Big) f_{,R} \right.\\
& \left.- \frac{1}{2}g_{\mu\nu}f \right]\mathrm{\delta g^{\mu\nu}} - \oint_{\partial \mathcal{V}} \mathrm{d^3x} \ \sqrt{h}f_{,R} h^{\mu\nu}\partial_\sigma(\delta g_{\mu\nu})n^\sigma
\end{split}\ee
where, comma ($,$) stands for ordinary derivative, $ \nabla_\mu $ stands for usual covariant derivative, $ \Box = g^{\mu\nu}\nabla_\mu\nabla_\nu$, $h$ is the determinant of the induced metric $h_{ij}$ and $n_{\mu}$ is the unit normal to the hypersurface. The boundary term in the above expression neither can be set to vanish at the boundary nor can be expressed in terms of standard surface invariants as Gibbons-Hawking-York term for general theory of relativity(GTR), following the only consideration that $\delta g_{\mu\nu}\Big{|}_{\partial\mathcal{V}} = 0$ \cite{madsen}. However, the boundary term appearing in equation (\ref{c5}) may be expressed in terms of standard surface invariants as \cite{barth, madsen, nakamura, dyer}
\bes\label{bt} \delta [{\big(K f_{,R}\big)} d\Sigma ]= d\Sigma
\left[K f_{,RR}\delta R + \frac{f_{,R}}{2}h^{\mu\nu}\partial_\sigma(\delta g_{\nu\mu})n^\sigma\right]
\end{split}\ee
where $d\Sigma = \oint_{\partial {\mathcal V}}\mathrm{d^3x} \sqrt{h}$. Clearly, the right hand side of (\ref{bt}) produces the above boundary term appearing in equation (2), provided $\delta R\Big{|}_{\partial\mathcal V} = 0$. Field equations for $f(R)$ theory of gravity are then obtained as usual. Hence, the complete action for $f(R)$ theory of gravity is expressed as
\be \label{c7} A = \int_{\mathcal{V}}\sqrt{-g}f(R)\mathrm{d^4x} + 2\oint_{\partial\mathcal{V}}\sqrt{h}f'(R)K \  \mathrm{d^3x},\ee
keeping in mind that the condition $R = \text{constant}$ has been set at the boundary. Such a restriction for higher order theory of gravity has been a debatable issue over years. The debate essentially stems from the fact that there is no convincing physics for $\delta R\Big{|}_{\partial\mathcal V} = 0$. The reason is, if $R$ is kept fixed on every space-like hypersurface, then the only solution admissible to the corresponding field equations is de-Sitter or anti de-Sitter(dS/AdS), depending on the signature of the constant \cite{Ellis}. While GTR admits indefinitely different types of solutions corresponding to different choice of energy-momentum tensor, higher order theory of gravity admits only a unique solution, appeared to be untenable. This initiated to develop even a new variation principle - the Palatini variational principle \cite{Palatini}, in which the action is varied both with respect to the metric tensor $g_{\mu\nu}$ and the connection $\Gamma^{\lambda}{_{\mu\nu}}$, which are treated as independent variables. This technique does not produce a boundary term under variation and therefore it is not required to make any restriction. Nevertheless, in the process it does not also produce Gibbons-Hawking-York boundary term for GTR also, and as a result, one looses most cherished concept of black hole entropy. Under this situation, our present aim is to show - ``the fact that higher order theory of gravity does not admit any solution other than dS/AdS in metric variational technique, is supported by Noether symmetry consideration". Noether symmetry had been applied initially by Rugeiro and his co-workers \cite{Rugeiro} in cosmology, to find the form of the scalar potential in scalar-tensor theory of gravity. The potential so found was exponential, which is suitable to drive inflation in the early stage of cosmic evolution. Till date, there has been numerous attempts in this field, in which its application on scalar-tensor theory of gravity \cite{c-a} are of particular importance. It has also been applied by several authors in the context of higher order theory of gravity \cite{higher1, higher2, higher3, higher4}, overlooking an important issue that we shall explore here.\\

\noindent
First, we take an action in the form
\be\label{a} A = \int\sqrt{-g}d^4x f(R) + \Sigma \ee
where, $\Sigma$ is the boundary term for $f(R)$ given in (\ref{c7}).
To express the above action in canonical form in the Robertson-Walker minisuperspace
\be ds^2 = -dt^2 + a^2 \left[{dr^2\over {1-kr^2}} + r^2(d\theta^2 + \sin^2\theta d\phi^2)\right]\ee
say, it is customary to treat $R-6\Big({\ddot a\over a} + {\dot a^2\over a^2} + {k\over a^2}\Big) =0$ as a constraint and introduce it through a Lagrange multiplier $\lambda$ into the action (\ref{a}) as \cite{higher2}
\bes A = \int\sqrt{-g}d^4x \Big[f(R) - \lambda\Big\{R-6\Big({\ddot a\over a} + {\dot a^2\over a^2} + {k\over a^2}\Big)\Big\}\Big] + \Sigma.\end{split}\ee
In the process, $a$ and $R$ are treated as canonical variables by expressing the action as $A = \int \mathcal{L}(a, \dot a, R, \dot R) dt$. Mixing up of minisuperspace and superspace variables might appear strange, since the Ricci scalar $R$ is not independent of $a$ and $\dot a$. But we know that canonical formulation of higher order theory of gravity requires the introduction of auxiliary variable, which might be different in the case the form of $f(R)$ is known a-priori. However, for arbitrary form of $f(R)$, this is the only way to obtain canonical formulation. In fact, the definition of $R$ in terms of ($a, \dot a, \ddot a$) introduces a constraint which eliminates the second-and higher-order derivatives in (7), so that a system of second-order differential equations in ($a, R$) is realized. Thus the Lagrange multiplier has not been introduced in an ad-hoc manner, since it is related to the symmetries and conservation laws. Now varying the action with respect to $R$ one gets $\lambda = f_{,R}$. Substituting it in the action and performing integration by parts, the boundary term $\Sigma$ gets cancelled and one is left with the following canonical action
\bes A = \int\Big[ a^3(f-R f_{,R}) -6a(\dot a^2-k) f_{,R} -6a^2 \dot a\dot R f_{,RR}\Big]dt.\end{split},\ee
where, comma stands for derivative. Now, in order to find an appropriate form of $f(R)$, let us impose Noether symmetry $\pounds_X L = XL =0$, where $\pounds$ stands for Lie derivative and $X$ is the vector field. Then equating the coefficients of $\dot a^2$, $\dot R^2$, $\dot R\dot a$ and the terms independent of these separately equal to zero, one obtains, following set of equations, viz.,
\begin{subequations}\begin{align}
& a(a\beta)_{,a}f_{,RR} + (\alpha + 2a\alpha_{,a})f_{,R} = 0 \label{a1}\\
& f_{,RR}\alpha_{,R} = 0 \label{a2}\\
& \beta a f_{,RRR}+(2\alpha + a\alpha_{,a}+a\beta_{,R})f_{,RR} + 2\alpha_{,R}f_{,R} = 0 \label{a3}\\
& \beta a(6k-a^2R)f_{,RR}+3\alpha[a^2(f-R f_{,R})+ 2kf_{,R}]= 0. \label{a4}
\end{align}\end{subequations}
Under the restriction $f_{,RR}\ne 0$, equations (\ref{a1}) through (\ref{a3}) lead to
\begin{subequations}\begin{align}
&\alpha = \alpha(a)\\
&{f_{,RRR}\over f_{,RR}}+{\beta_{2_{,R}}\over \beta_2} = {c_1\over \beta_2}\label{s1}\\
&\beta_2 {f_{,RR}\over f_{,R}} = {1\over c_2}\label{s2}\\
&\alpha_{,a} +2{\alpha\over a}=-c_1 \beta_1\label{s3}\\
&a^2\alpha_{,aa}+(3-2c_1c_2)a\alpha_{,a} - c_1c_2\alpha =0\label{s4}
\end{align}\end{subequations}
where, $\beta =\beta_1(a)\beta_2(R)$. Equations (\ref{s1}) and (\ref{s2}) then yield $c_1c_2 = 1$, and as a result equation (\ref{s4}) leads to
\be\label{s5} a^2\alpha_{,aa}+a\alpha_{,a} - \alpha =0 \ee
which may be solved immediately and hence $\beta_1$, in view of equation (\ref{s3}). However, $f(R)$ and $\beta_2(R)$ still remain arbitrary. So before solving equation (\ref{s5}) let us concentrate upon equation (\ref{a4}), which now takes the form
\be\label{s6} a^2(a\alpha_{,a}-\alpha)R f_{,R} -6k(a\alpha_{,a} +\alpha)f_{,R} +3a^2\alpha f =0.\ee
Now if $f(R)$, $f_{,R}$ and $R f_{,R}$ are all independent then the only solution to the above equation (\ref{s6}) is $\alpha =0$ and so is $\beta_1$, and therefore Noether symmetry does not exist. Therefore one has to explore Noether symmetry for particular form of $f(R)$.\\

\noindent
\textbf{Case-I $f(R) = R^n$.}\\
Equation (\ref{s6}) then reduces to
\be a^2[na\alpha_{,a}+(3-n)\alpha]R^n - 6k n (a\alpha_{,a}+\alpha)R^{n-1} = 0.\ee
clearly, setting coefficients to vanish, one ends up with $f(R) = R^{3\over 2}$ and all other equations are satisfied, leading to a conserved current ${d\over dt}(a\sqrt R)$. This is already a known result, which is also true in the matter dominated era \cite{higher2}. More importantly, this is the only possible form of $f(R)$, irrespective of arbitrary minimal or non-minimal coupling of matter \cite{higher3}. In particular, it was found that under a change of variable $z = a^2$, $z$ becomes cyclic for $f(R) = R^{3\over 2}$, in vacuum and in the presence of pressureless dust. However, the other option, stemming from equation (13) viz, $R = R_0$, where, $R_0$ is a constant, was overlooked earlier. Since, it is a solution to the field equations, so naturally all the Noether equations are satisfied. It might appear that the symmetry appearing from $R = R_0$ might be ignored. But we shall show below that this is the only solution of Noether symmetry for other forms of $f(R)$ and even in anisotropic models.\\

\noindent
\textbf{Case-II $f(R) = {R\over 16\pi G}+\gamma R^n$.}\\
Equation (\ref{s6}) now reads
\bes \gamma a^2[na\alpha_{,a}+(3-n)\alpha]R^n - 6\gamma k n (a\alpha_{,a}+\alpha)R^{n-1}\\& + {a^2\over 16\pi G}(a\alpha_{,a}+2\alpha)R -{3k\over 8\pi G}(a\alpha_{,a} + \alpha)= 0.\end{split}\ee
Clearly, the above equation is satisfied provided coefficients of different powers of $R$ vanish. However, in the process the second and the last term is solved for $\alpha$ as
\be \alpha = {\alpha_0 \over a},\ee
which also satisfies equation (\ref{s5}). For $n = {3\over 2}$, first term vanishes and one ends up with the condition $R = 0$. Nevertheless, for $n\ne {3\over 2}$, the above equation now reduces to
\be R^{(n-1)} = - {1\over 16\pi G \gamma}\Big[{(a\alpha_{,a}+2\alpha)\over na\alpha_{,a}+(3-n)\alpha}\Big]\ee
In view of the solution (15) above equation reads
\be R = \left[{1\over 16\pi G(2n - 3)}\right]^{1\over n-1},\ee
which is clearly a constant. One can solve for $\beta = {\beta_0\over a^2}$ to find the conserved current once again. However, it is not required, since $R = {\mathrm{constant}}$ is always a solution (dS/AdS) to higher order theory of gravity.\\

\noindent
\textbf{Case-III $f(R) = f_0 e^{nR}$.}\\
Equation (\ref{s6}) in this case is expressed as
\be a^2[a\alpha_{,a}-\alpha]n R - 6k n(a\alpha_{,a}+\alpha) + 3 a^2\alpha = 0\ee
which does not admit any solution other than $R = R_0$ - a constant. The above equation may then be solved to yield
\be \alpha = {\alpha_0\over a}, ~\textrm{provided},~~ R_0 = {3\over 2n}.\ee

\noindent
One can choose other forms of $f(R)$ and even couple it to some form of matter to end up with the same result that Noether symmetry only admits a constant Ricci scalar. However, situation is much more apparent while searching for Noether symmetry of $f(R)$ theory of gravity in anisotropic models.  As an example, let us take into account spatially symmetric Kantowski-Sachs, Bianchi-I and Bianchi-III metric,

\be ds^2 = -dt^2 + a^2 dr^2 + b^2[d\theta^2 + f_k^2 d\phi^2], \ee
where,
$$
f_k =
\left\{
  \begin{array}{lll}
    sin\theta & \Rightarrow k = + 1 & \text{for} \ \ (K-S) \\
    \theta & \Rightarrow k = 0 & \text{for} \ \ (B-I) \\
    sinh\theta & \Rightarrow k = - 1 & \text{for} \ \ (B-III)
  \end{array}
\right.
$$
and the Ricci scalar reads
\be R = 2\left(\frac{\ddot a}{a}+2\frac{\ddot b}{b}+2\frac{\dot a\dot b}{ab}+\frac{\dot b^2}{b^2} +\frac{k}{b^2} \right),\ee
so that the action
\be \begin{split} A = \int\Big[B f(R) - \lambda\Big\{R - 2\Big(\frac{\ddot a}{a}+2\frac{\ddot b}{b}+2\frac{\dot a\dot b}{ab}+\frac{\dot b^2}{b^2}\\
 +\frac{k}{b^2} \Big)\Big\}\Big]\sqrt{-g} dt +\Sigma,\end{split}\ee
is expressed as $A = \int\mathcal{L} (a, \dot a, b, \dot b, R, \dot R)dt$, treating $a$, $b$ and $R$ as canonical variables. The action is now varied with respect to $R$ to obtain $\lambda = Bf_{,R}$. Thus after removing total derivative terms which gets cancelled with $\Sigma$, the action takes the following canonical form,
\be \begin{split}
A &= \int\Big[ab^2(f - R f_{,R})-2f_{,R}(2b\dot a\dot b + a\dot b^2 - k a)\\
& - 2 f_{,RR}(b^2\dot a + 2ab\dot b)\dot R\Big]dt .\end{split}\ee
At this stage, imposing Noether symmetry ($\pounds_X L = X L = 0$), following set of equation are found
\begin{subequations}\begin{align}
& \frac{f_{,RR}}{f_{,R}} = -\frac{2\beta_{,a}}{b\gamma_{,a}} \label{b1}\\
& \frac{f_{,RR}}{f_{,R}} = -\frac{\alpha+2b\alpha_{,b}+2a\beta_{,b}}{a(\gamma+2b\gamma_{,b})} \label{b2}\\
& \frac{f_{,RR}}{f_{,R}} = -2\frac{b\alpha_{,a}+\beta+a\beta_{,a}+b\beta_{,b}}{b(2\gamma+2a\gamma_{,a}+b\gamma_{,b})} \\
& b\alpha_{,R} + 2a \beta_{,R} = 0 \label{b3} \\
& \frac{f_{,RRR}}{f_{,R}}+\frac{f_{,RR}}{f_{,R}}\Big[\frac{2b\alpha+b^2\alpha_{,b}+2a\beta+2ab\beta_{,b} +2ab\gamma_{,R}}{2ab\gamma}\Big] \nonumber \\
& +\frac{b\alpha_{,R}+a\beta_{,R}}{ab\gamma}=0 \\
& \frac{f_{,RRR}}{f_{,R}}+\frac{f_{,RR}}{f_{,R}}\left[\frac{2\beta+b\alpha_{,a}+2a\beta_{,a}+b\gamma_{,R}}{b\gamma}\right]
   +2\frac{\beta_{,R}}{b\gamma}=0. \\
& f_{,RR}+\left[\frac{b^2R\alpha-2k\alpha+2abR\beta}{a(b^2 R - 2k)\gamma}\right]f_{,R} \nonumber \\
& -\left[\frac{b(b\alpha+2a\beta)}{a(b^2 R -2k)\gamma}\right]f =0 \label{b}
\end{align}\end{subequations}
Now under the assumption that $f(R)$ is non-linear in $R$ and expressing $\alpha$, $\beta$ and $\gamma$ in the following form,
\be \begin{split}
\alpha &= A(a, b)D_1(R) \\
\beta &= B(a, b)D_2(R) \\
\gamma &= C(a,b)D_3(R)
\end{split}\ee
one can deduce the following relations
\be \frac{D_3f_{,RR}}{D_2f_{,R}} = n_1 \ \ \text{and} \ \ D_2 = m_1D_1, \ee
in view of equations (\ref{b1}) and (\ref{b2}). Again using (\ref{b3}) one obtains the following relation,
\be b A + 2m_1aB = 0 .\label{b4}\ee
Now in view of equations (\ref{b}) and (\ref{b4}) one finally ends up with,
\be \left[m_1 n_1 - \frac{2k A}{a(b^2 R - 2k)C}\right]f_{,R} = 0,\ee
implying
\be R = \frac{2k}{b^2}\left[\frac{A}{m_1 n_1 a C} + 1\right] = n_2.\ee
where $n_2$ is a separation constant. Thus, $R$ turns out to be a constant. Now using (\ref{b}) one can only relate $\alpha$ and $\beta$ as $\alpha = -\frac{2a}{b}\beta$, while $\gamma$ remains arbitrary. This solution was found earlier \cite{higher4}, but was overruled since Birkhoff's theorem that Schwarzschild's solution is the unique spherically symmetric vacuum solution does not hold for $f(R)$ theory of gravity. However, $f(R)$ theory of gravity has got no obligation to Birkhoff's theorem. Earlier \cite{higher4} we wondered why despite the fact that there are indefinitely large number of curvature invariants together with coupling parameters corresponding to scalar field, Noether symmetry does not exist both in isotropic and anisotropic models, except for the very special one ($f(R) \propto R^{3\over 2}$) obtained in isotropic case \cite{higher2, higher3, higher4}. Now it is clear that, earlier we have overruled dS/AdS solutions. It is noteworthy that a different canonical formulation of $R^2$ theory of gravity through auxiliary variable also yields the only solution $R =$constant as a result of Noether symmetry \cite{kna}. The fact that Noether symmetry only admits $R = \mathrm{constant}$ solution, appears to administer the condition $\delta R\Big{|}_{\partial{\mathcal V}} = 0$, required to validate metric variational method.\\

\noindent
In recent years $f(R)$ theory of gravity along with certain extended version of it have been advocated as alternative to the dark energy. Nothing could have been better if geometry really plays the role of dark energy. Nevertheless, there are several reasons to defy such claim. First of all, higher order terms considered for the purpose stated, are not distinguished at all, since neither these are generated under one-loop quantum gravitational correction nor from any other physical consequence, like Noether symmetry. Further, all the results viz. the bridge between early inflation and late-time cosmic acceleration, to pass through solar test etc. which emerged from $f(R)$ theory of gravity resulted from scalar-tensor equivalence (Einstein's and Jordan's frame). No-body, so far paid any heed in the issue of the associated boundary term, which does not admit solutions other than dS/AdS. The situation thus stands is, either one should rule-out the technique adopted for canonical formulation of $f(R)$ theory by treating both the superspace and minisuperspace variables together. In that case there is no answer to the issue of boundary term unless one discards metric variational technique and adopts Palatini formulation. This is a good option, since equivalence principle is not supposed to be valid in higher order theory of gravity. However, in the process one has to sacrifice one of the most cherished concepts of gravitational physics, viz., the Black-Hole entropy and also has to rely upon the solutions obtained under scalar tensor equivalence, which has no quantum analogue. On the other hand, one should stick to the metric variational principle. In that case, canonical formulation of $f(R)$ theory of gravity is possible only under the introduction of Lagrange multiplier which is anholonomic constraint being capable of reducing the dynamics. It is also related to the existence of Noether symmetries, which helps to extract exact solutions. The Lagrange multiplier approach also helps in the formulation of covariant renormalizable gravity \cite{lagrange}. In recent years, the Lagrange multiplier technique has been found to play much important role in the context of quantum cosmology and higher order theories of gravity. For example, it has been shown that the existence of  Noether symmetries implies a subset of the general solution of the Wheeler-DeWitt equation where the oscillating behaviors are selected naturally. In the process, the Hartle criterion is related to Noether symmetry and hence to the classical trajectories \cite{lagrange1}. In connection with $f(R)$ theory of gravity, Lagrange multiplier has been found to play the role of cosmological constant and inflationary behavior is asymptotically recovered \cite{lagrange1}. It has also been demonstrated that, using Lagrange multiplier in connection with Gauss-Bonnet-dilatonic coupling, one gets large number of accelerating cosmological models, including the phantom ones where the dilatonic kinetic term is canonical. In particular, the Lagrange multiplier behaves as a sort of dust fluid that realizes the transitions between matter-dominated and dark energy epochs \cite{lagrange2}. All these results favour the use of Lagrange multiplier technique for canonical formulation of $f(R)$ theory of gravity. But then, the technique although validates the boundary condition suitably, nevertheless, the resulting solutions are not suitable to explain late-time cosmic acceleration. This is because, such solutions do not admit early long decelerated era, required for structure formation and also do not validate Birkhoff's theorem. It is also important to note that the resulting dS/AdS solutions are necessary and also well-behaved from the standpoint of Inflation required in the early universe, since graceful exit is possible under reheating, for example following particle creation, giving way to the hot Big-Bang era. Hence it appears that one should rule-out $f(R)$ theory of gravity and all of its extended versions, as viable candidates for dark energy.

\end{document}